\documentclass[conference]{IEEEtran}
\IEEEoverridecommandlockouts
\usepackage{cite}
\usepackage{amsmath,amssymb,amsfonts}
\usepackage{algorithmic}
\usepackage{graphicx}
\usepackage{textcomp}
\usepackage{xcolor}
\def\BibTeX{{\rm B\kern-.05em{\sc i\kern-.025em b}\kern-.08em
    T\kern-.1667em\lower.7ex\hbox{E}\kern-.125emX}}

\usepackage{CJKutf8}
\usepackage{booktabs} 
\usepackage{tabularx}
\usepackage{array}
\usepackage{multirow}
\usepackage{amsmath}
\usepackage{amssymb}
\usepackage{color}
\usepackage{bbm}
\usepackage{bbding}
\usepackage{pifont} 
\definecolor{rev}{rgb}{0.,0.,0.}
\renewcommand{\vec}[1]{\boldsymbol{#1}}
\makeatletter
\renewcommand{\@makefnmark}{}  % 不显示角标
\makeatother

\begin{document}

\title{BrainCognizer: Brain Decoding with Human Visual Cognition Simulation for fMRI-to-Image Reconstruction
}

\author{
% Anonymized Authors
\IEEEauthorblockN{Guoying Sun$^{1}\ \ \ $ Weiyu Guo$^{2}\ \ \ $   Tong Shao$^{1}\ \ \ $  Yang Yang$^{1}\ \ \ $  Haijin Zeng$^{3}\ \ \ $  Jie Liu$^{1}\ \ \ $  Jingyong Su$^{1*}$\thanks{* Corresponding author, email-to: sujingyong@hit.edu.cn}} 
\IEEEauthorblockA{1. School of Computer Science and Technology, Harbin Institute of Technology (Shenzhen), China \\
2. Thrust of Artificial Intelligence, The Hong Kong University of Science and Technology (Guangzhou), China\\
3. Department of Psychiatry, Harvard University, USA} \\
% email address or ORCID
% \and
% \IEEEauthorblockN{Weiyu Guo$^{2}$}
% % \IEEEauthorblockA{\textit{dept. name of organization (of Aff.)} \\
% % \textit{name of organization (of Aff.)}\\
% % City, Country \\
% % email address or ORCID}
% \and
% \IEEEauthorblockN{Tong Shao$^{1}$}
% % \IEEEauthorblockA{\textit{dept. name of organization (of Aff.)} \\
% % \textit{name of organization (of Aff.)}\\
% % City, Country \\
% % email address or ORCID}
% \and
% \IEEEauthorblockN{Yang Yang$^{1}$}
% % \IEEEauthorblockA{\textit{dept. name of organization (of Aff.)} \\
% % \textit{name of organization (of Aff.)}\\
% % City, Country \\
% % email address or ORCID}
% \and
% \IEEEauthorblockN{Haijin Zeng$^{3}$}
% % \IEEEauthorblockA{\textit{dept. name of organization (of Aff.)} \\
% % \textit{name of organization (of Aff.)}\\
% % City, Country \\
% % email address or ORCID}
% \and
% \IEEEauthorblockN{Jie Liu$^{1}$}
% % \IEEEauthorblockA{\textit{dept. name of organization (of Aff.)} \\
% % \textit{name of organization (of Aff.)}\\
% % City, Country \\
% % email address or ORCID}
% \IEEEauthorblockN{Jingyong Su$^{1*}$}

}

\maketitle

\begin{abstract}
Brain decoding is a key neuroscience field that reconstructs the visual stimuli from brain activity with fMRI, which helps illuminate how the brain represents the world. fMRI-to-image reconstruction has achieved impressive progress by leveraging diffusion models. However, brain signals infused with prior knowledge and associations exhibit a significant information asymmetry when compared to raw visual features, still posing challenges for decoding fMRI representations under the supervision of images. Consequently, the reconstructed images often lack fine-grained visual fidelity, such as missing attributes and distorted spatial relationships. To tackle this challenge, we propose BrainCognizer, a novel brain decoding model inspired by human visual cognition, which explores multi-level semantics and correlations without fine-tuning of generative models. Specifically, BrainCognizer introduces two modules: the Cognitive Integration Module which incorporates prior human knowledge to extract hierarchical region semantics; and the Cognitive Correlation Module which captures contextual semantic relationships across regions. Incorporating these two modules enhances intra-region semantic consistency and maintains inter-region contextual associations, thereby facilitating fine-grained brain decoding. Moreover, we quantitatively interpret our components from a neuroscience perspective and analyze the associations between different visual patterns and brain functions. Extensive quantitative and qualitative experiments demonstrate that BrainCognizer outperforms state-of-the-art approaches on multiple evaluation metrics. 
% The code and models will be made publicly available.
Our code is released publicly at https://github.com/Grace160/BrainCognizer.

% Specifically, two components mimic the processes within the human visual cognition system: the Cognitive Integration Module (CIM) which extracts hierarchical semantic representations from fMRI data to integrate region-level information for holistic understanding; and the Cognitive Correlation Module (CCM) which captures high-level semantic relationships to model the perception of semantic correlations for conceptual comprehension.
% Incorporating cognitively inspired features into the feature extraction process introduces intra-region semantic consistency and inter-region correlations, thereby facilitating fine-grained brain decoding.
\end{abstract}


\section{Introduction}
\label{sec:introduction}
A challenging goal in neuroscience is to decode information from brain activity. \footnotetext{This paper was accepted by IEEE International Conference on Bioinformatics and Biomedicine (BIBM) 2025.}
The functional magnetic resonance imaging (fMRI) signals with changes in brain oxygen levels reflect the brain activity in response to external stimuli~\cite{ogawa1990oxygenation}.
% fMRI-to-image reconstruction~\cite{brainactiv,guo2025surveyfmriimagereconstruction,brainsail} is an emerging task in brain decoding, which aims to reconstruct images from fMRI elicited by visual stimuli, establishing the connection between brain activity and vision. 
fMRI-to-image reconstruction~\cite{brainactiv,guo2025surveyfmriimagereconstruction,brainsail} aims to reconstruct visual stimuli from fMRI signals, bridging brain activity and perception.
{\color{rev}
% However, decoding visual information from fMRI is limited by the scarcity of fMRI-image pairs~\cite{beliy2019voxels,xia2024dream} and the low Signal-to-Noise Ratio~\cite{comby2023denoising} caused by attentional distractions and associative interferences during data collection. Numerous pixel-to-voxel and generative adversarial network-based methods~\cite{st2018generative,seeliger2018generative,lin2019dcnn,ren2021reconstructing} tend to generate blurry images with poor semantic consistency, limiting their effectiveness in fMRI-to-image reconstruction.
Due to the scarcity of fMRI-image pairs~\cite{beliy2019voxels,xia2024dream} and the inherently low signal-to-noise ratio~\cite{comby2023denoising} of fMRI caused by attentional distractions and associative interferences during data collection, traditional voxel-to-pixel methods~\cite{st2018generative,seeliger2018generative,lin2019dcnn,ren2021reconstructing} often generate blurry images with poor semantic consistency.

% Benefiting from the pre-trained cross-modal model CLIP~\cite{radford2021learning}, which provides an effective representation space with structural similarities to the human brain~\cite{yang2024brain,wang2023better} and exhibits superior explanatory power for cortical activity~\cite{zhouclip}, as well as latent diffusion models~\cite{ddpm} with high-fidelity visual reconstruction capabilities, fMRI-to-image reconstruction has achieved notable progress.
Recently, the emergence of cross-modal foundation models such as CLIP~\cite{radford2021learning}, alongside the rapid advances in generative techniques like diffusion models~\cite{ddpm,ldm,vd}, has led to substantial breakthroughs in fMRI-to-image reconstruction.
The current pipeline of fMRI-to-image reconstruction~\cite{ozcelik2023natural,scotti2024reconstructing,takagi2023high,wang2024mindbridge} employs an encoder to learn neural representations from fMRI signals, which are aligned with features extracted by images. 
Features reflecting visual information decoded from brain activity serve as semantic conditions for diffusion models, which leverages rich priors and strong generative capabilities to produce high-fidelity image reconstructions.
Therefore, the fMRI-to-image task is simplified to aligning fMRI-derived features with image representations obtained from foundation models, enabling the exploration of their encoding capabilities.

\begin{figure}[tb]
  \centering
  \includegraphics[height=5.5cm]{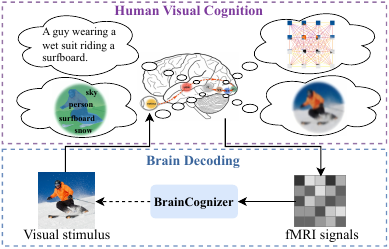}
  \caption{ 
  % Schematic diagram of the brain signal generation process in neuroscience. Low-level cortical areas respond to edge and color information, while high-level cortical areas capture higher-level semantic cognitive and relationship understanding.
% Overview of brain signal generation and decoding: fMRI signals reflect low-level encoding of edges and colors, and high-level extraction of semantics and relationships.
  The brain decoding process by simulating the human visual cognition.
  BrainCognizer extracts (1) coarse-grained semantics from captions, (2) fine-grained visual prior knowledge, (3) regional contextual relationships, and (4) low-level visual features for fMRI decoding and reconstruction.
  }
  \label{fig:head}
\end{figure}

However, fMRI representation decoding under the direct supervision of CLIP features is often prone to overfitting~\cite{luanimate}, which we believe is essentially caused by dimension discrepancy.
% However, due to the low signal-to-noise ratio of fMRI, directly learning a mapping between fMRI and the semantic condition is prone to overfitting~\cite{luanimate}, which we believe is essentially caused by dimension discrepancy.
% The fMRI signals encompass complex and diverse neural representations with \textit{\textbf{high-dimensional}} semantic information~\cite{gazzaniga2009cognitive}, where low-level cortical areas respond to edge and color information and high-level cortical areas capture higher-level cognitive functions influencing perception, together forming a coherent perceptual experience, as shown in Fig. \ref{fig:head}. 
fMRI signals encode complex and diverse neural representations, encompassing both low-level visual perceptions (such as color, shape~\cite{felleman1991distributed}, and layout~\cite{grill2014functional,kravitz2011new}) and high-level cognitive features (including semantic categories, memory associations, and subjective experiences~\cite{huth2012continuous}). In contrast to \textit{low-dimensional} CLIP features which are manually designed and primarily capture visual categories, fMRI signals serve as extremely \textit{high-dimensional} carriers of rich brain information.
% The fMRI signals encompass complex and diverse neural representations with \textit{high-dimensional} semantic information~\cite{gazzaniga2009cognitive}, where the interplay between low-level and high-level cortical areas forms a coherent perceptual experience, as shown in Fig.~\ref{fig:head}. 
% Specifically, early visual areas extract low- and mid-level features such as edges, color, and shape~\cite{felleman1991distributed}, while high-level visual areas capture more conceptual representations, including object semantics, scene structure, and spatial relations~\cite{grill2014functional,kravitz2011new}.
% The coordinated activity across these regions enables efficient understanding of complex visual scenes, and generates the complex brain signals observed in fMRI.

% Consequently, this significant discrepancy in dimensionality leads to an \textit{information asymmetry}, making it challenging to decode the necessary semantics from the complex fMRI signals for effectively guiding diffusion.
Consequently, this significant discrepancy in dimensionality leads to an \textit{information asymmetry}, hindering effective semantic decoding from complex fMRI signals to guide diffusion.
% Considering the \textit{low-dimensional} features with limited semantics \cite{luanimate} of CLIP, the intrinsic information bottleneck leads to \textit{information asymmetry} between brain signals containing prior knowledge and visual features, posing challenges for accurate brain decoding.
Existing methods~\cite{scotti2024reconstructing,lu2023minddiffuser,wang2024mindbridge} fail to address the representational gap and ignore the inherent information asymmetry, thereby impeding the reconstruction of fine-grained details, including semantic attributes, spatial relationships, and contextual information, which are shown in Fig.~\ref{fig:headrec}.
% attributes such as snow, umbrella, and horse are missing, and the contextual and spatial structures appear disordered.

To tackle the challenge, we propose a novel brain decoding model, \textbf{BrainCognizer}, which enhances the hierarchical semantics and maintains region contextual correlations through the simulation of human visual cognition.
As illustrated in Fig.~\ref{fig:head}, BrainCognizer enhances the fMRI decoding by leveraging prior knowledge derived from human visual cognition, effectively mitigating the reconstruction quality degradation caused by information asymmetry.

Specifically, we first propose the \textit{Cognitive Integration Module (CIM)} to enhance the diversity of hierarchical region semantics by simulating the human visual holistic perception~\cite{koffka1922perception}.
Subsequently, we propose the \textit{Cognitive Correlation Module (CCM)}, which captures inter-region semantic differences and contextual relationships by topological correlation, simulating the brain differential perception~\cite{masland2012neuronal,poggio1990network}.
% Incorporating these modules into brain decoding enhances fMRI representation and harmonizes inter-modality disparities. 
% Our framework integrates CLIP and diffusion models to reconstruct high-fidelity semantic images without fine-tuning of generative models.
BrainCognizer leverages prior knowledge from human cognition to skillfully uncover semantics that guide the diffusion effectively, without any fine-tuning of the generative model.

Furthermore, we perform neuroscientific analyses of the embedded semantic and relational representations, thereby enhancing functional interpretability from a cognitive perspective.
In summary, our contributions are as follows: 

\textbf{1)} We propose BrainCognizer, a fine-grained brain decoding framework that mitigates information asymmetry by exploring hierarchical semantics and region contextual relations without fine-tuning of generative models.

\textbf{2)} We propose CIM and CCM inspired by human visual cognition, and quantitatively interpret them via brain function visualization, revealing that our method explores more fMRI voxels for decoding.

\textbf{3)} Extensive experimental results demonstrate the qualitative and quantitative fidelity of BrainCognizer, which achieves state-of-the-art performance on multiple evaluation metrics.

\begin{figure}[tb]
\centering
  \includegraphics[height=7.2cm]{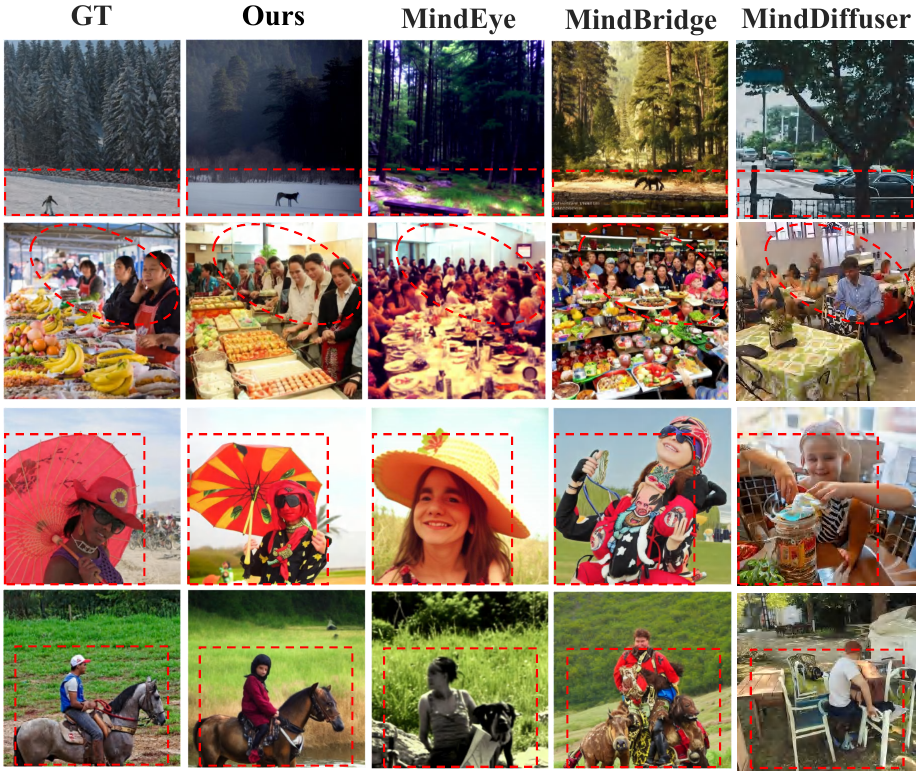}
  \caption{
  % Comparison of image reconstruction results.
  Image reconstruction examples.
  % Highlighted regions reveal missing attributes (rows 1, 3, 4) and spatial/contextual errors (rows 2–4).
  % Highlighted regions exhibit missing attributes and spatial or contextual disorder.
  The highlighted regions indicate comparison.
  Existing methods fail to reconstruct fine-grained semantic attributes, such as the snow in the 1st row or the umbrella and sunglasses in the 3rd row. They also fail to recover contextual layouts, such as the people in the 2nd row.
  Our BrainCognizer effectively mitigates these issues.
  }
  \label{fig:headrec}
\end{figure}

\begin{figure*}[tb]
  \centering
  \includegraphics[height=6.9cm]{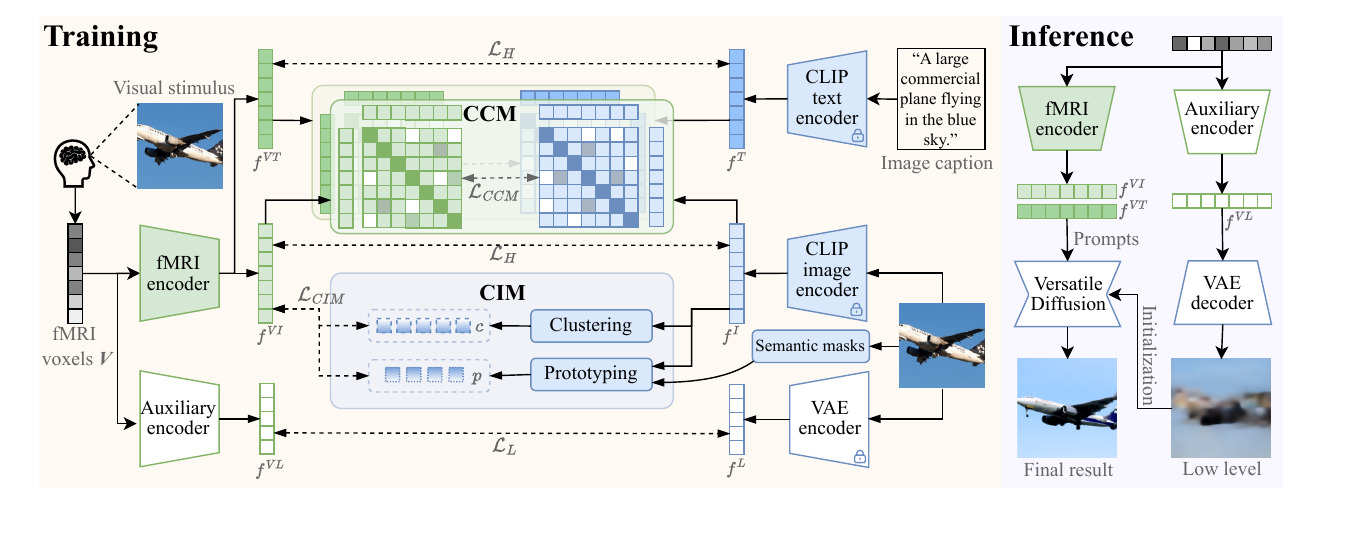}
  \caption{The framework of BrainCognizer. CIM and CCM simulate human visual cognition to extract intrinsic information, guiding fMRI for more comprehensive representations. During inference, the outputs of the auxiliary encoder initialize the diffusion process and subsequently the fMRI features serve as prompts to constrain the reconstruction of the visual stimulus.
  }
  \label{fig:model}
\end{figure*}

\section{Related Work}
{\color{rev}
% \subsection{Visual Neuroscience.}
% Reconstructing visual stimuli from brain signals requires understanding how the brain processes and interprets visual stimuli. The Gestalt principles~\cite{koffka1922perception} in neuroscience reveal the brain’s inherent tendency to organize visual information into regional coherent wholes, rather than processing discrete, independent elements in isolation. For example, the visual system tends to perceive a person as a whole, rather than as separate parts such as hands, arms, or legs. The concept of differential perception~\cite{masland2012neuronal,poggio1990network} in neuroscience shows that the brain processes visual information by comparing and distinguishing the relative differences among various stimuli. The Gestalt Principle and differential perception emphasize that perception is not merely the simple addition of individual stimuli, but a holistic and relative process that together forms the complex mechanisms of human perception. Inspired by the above visual perception principles, we simulate the human visual perception process to assist in fMRI-to-image reconstruction.

\subsection{Visual Stimulus Decoding from fMRI.}

Previous methods~\cite{brainclip,lin2022mind,rpi,ren2021reconstructing,takagi2023high,ozcelik2023natural,xia2024dream,ozcelik2022reconstruction} decode fMRI features directly supervised by CLIP features, and use them to guide the diffusion.
However, the large modality gap between fMRI and CLIP features, combined with the limited amount of annotated data, leads to overfitting.

Methods~\cite{xia2024dream,beliy2019voxels} attempt to alleviate it by leveraging large-scale unimodal or unlabeled data through self-supervised learning.
% MindBridge~\cite{wang2024mindbridge} reduces the modality gap by leveraging data from multiple subjects to improve individual-level generalization.
MindBridge~\cite{wang2024mindbridge} narrows the modality gap by leveraging multi-subject data to enhance individual generalization.
Nevertheless, their effectiveness is hindered by inter-subject variability. 
% Huo et al.~\cite{neuropictor} reduce the modality gap by transforming fMRI signals into 2D representations like images and introducing new encoders such as vision transformer. However, they increase training overhead and compromise the brain-region priors inherent in voxel-wise fMRI data due to the naive format conversion.
% In addition, MindEye~\cite{scotti2024reconstructing} and CLIP-MUSED~\cite{zhouclip} improve the encoder architecture to enhance the representational capacity of fMRI decoding.
% MindDiffuser~\cite{lu2023minddiffuser} further enhances image fidelity through post-processing with semantic constraints on the generated images.
MindDiffuser~\cite{lu2023minddiffuser} further enhances image fidelity via post-processing with semantic constraints.
In addition, MindEye~\cite{scotti2024reconstructing} improves the encoder architecture to enhance the representational capacity of fMRI decoding.

These methods overlook the fact that the fundamental cause of the modality gap lies in the information asymmetry between fMRI signals with rich prior knowledge and CLIP visual features.
To address this issue, we propose BrainCognizer, which enhances the fMRI decoding by leveraging prior knowledge from human visual cognition, mitigating the reconstruction quality degradation caused by information asymmetry.

\subsection{Foundation Models for Visual Decoding.}
Generation models have been widely applied in visual reconstruction. GAN-based methods~\cite{st2018generative,lin2019dcnn,ren2021reconstructing,gu2024decoding,lin2022mind} generate blurry images that lack semantic content, whereas recent studies~\cite{takagi2023high,ozcelik2023natural,scotti2024reconstructing,wang2024mindbridge} show that diffusion significantly improves both image quality and semantic fidelity.

fMRI signals are decoded under the supervision of CLIP features, and then used to guide the diffusion generation.
CLIP~\cite{radford2021learning} provides an effective representation space with structural similarities to the human brain~\cite{yang2024brain,wang2023better} and exhibits superior explanatory power for cortical activity~\cite{zhouclip}, significantly facilitating fMRI decoding.
To leverage more comprehensive CLIP semantics, Versatile Diffusion model~\cite{vd} conditions the generation process on both text and image features of CLIP, thereby improving image fidelity.

Furthermore, the VAE~\cite{vae} in the diffusion captures low-level visual codebooks. By mapping fMRI signals into the VAE latent space, the reconstruction process starts from a semantically blur prior instead of random noise, thus mitigating the stochasticity introduced by diffusion and improving reconstruction fidelity.

% LDMs are initialized by the latent space of variational autoencoders (VAE) and guided by semantic conditions of CLIP-shared space.
% The Stable Diffusion model~\cite{ldm} uses text captions for semantic control, while the Versatile Diffusion model~\cite{vd} further incorporates image-based semantic guidance to enhance fidelity.
% Other visual properties, such as sketches and depth maps, are challenging to decode directly from brain signals, which can introduce additional noise.
% For semantic fidelity and controllability in visual reconstruction, we adopt the versatile diffusion model to incorporate both textual and visual information as semantic conditions, thereby enhancing the reliability of reconstructed images.

% Low-level visual properties such as sketches and depth maps are difficult to decode directly from brain signals, potentially introducing noise. 
% Some diffusion models~\cite{ldm} support low-level visual properties such as sketches and depth maps to guide image generation, but these are difficult to decode directly from brain signals and may introduce noise.
% To enhance semantic fidelity and controllability, we adopt the Versatile Diffusion, which incorporates both textual and visual cues as semantic conditions to improve reconstruction reliability.
}

\section{Method}
Inspired by the human visual cognition, we propose a novel fMRI-to-image reconstruction method, BrainCognizer, to improve the inadequate characterization caused by the information asymmetry between fMRI and images. 
BrainCognizer mainly comprises the Cognitive Integration Module (CIM) for intra-region semantic feature aggregation and the Cognitive Correlation Module (CCM) for inter-region topological correlations. The overview of BrainCognizer is illustrated in Fig.~\ref{fig:model}.

\subsection{Pipeline}
\label{sec:lowlevel-decoding}
% {\color{rev}Given a dataset \( D = \{(V_i, I_i)\}_{i=1}^N \) of brain fMRI voxels and corresponding visual images, the goal is to reconstruct \( I_i \) from \( V_i \) by learning a mapping from brain activity and visual content. Benefiting from the LDM's powerful image generation capabilities, this mapping is simplified to align the embeddings of fMRI voxel and LDM for learning the encoding capabilities of foundation models. 

% To reconstruct visual images from fMRI data, we build our pipeline upon the Versatile Diffusion (VD) model~\cite{vd} as shown in Fig.~\ref{fig:model}. VD is initialized from the latent space of VAE and guided by CLIP embeddings as semantic conditions to generate images. Accordingly, we use the VAE latent features $\vec{f}^L$, CLIP image features $\vec{f}^I$, and CLIP text features $\vec{f}^T$, extracted from the stimulus image and its caption, as ground-truth supervision. 
% These features capture low-level visual details and multi-granularity high-level semantics, providing effective guidance for fMRI-to-image reconstruction.

% To 从fMRI中解码更多pattern, we design a novel pipeline for fmri-to-image reconstruction. It decodes the low-level details 借助 VAE and 多层次语义 借助CLIP from fMRI 并借助Versatile Diffusion (VD) model~\cite{vd}实现视觉图像重建。
To decode richer visual patterns from fMRI, we propose a novel pipeline for fMRI-to-image reconstruction. 
It extracts low-level visual details through the VAE and decodes multi-level semantic representations via CLIP, while utilizing the Versatile Diffusion (VD)~\cite{vd} to reconstruct visual images.
VD takes the low-level results as the initial latent of VAE and generates the final reconstruction guided by the conditions of multi-level semantic representations.

% To enhance fMRI decoding, we integrate the semantic priors of CLIP and the generative capacities of diffusion models in feature learning.
For decoding low-level visual attributes, such as spatial structure and color composition, we introduce a lightweight auxiliary encoder composed of CNN and MLP. It decodes low-level features $\vec{f}^{VL} \in \mathbb{R}^{4\times64\times64}$ from fMRI and aligns them with the latents $f_{L}$ of VAE by mean absolute error loss, enabling the low-level blur perception from brain activity:
\begin{equation}
    % \mathcal{L}_{lowlevel} = \frac{1}{B}\sum^{B}_{b=1}|\vec{f}^{L}_{b}-\vec{f}^{VL}_{b}|.
    \mathcal{L}_{L} = |\vec{f}^{L}-\vec{f}^{VL}|.
\end{equation}

For the decoding of high-level semantic information, we incorporate a dual-head fMRI encoder to extract features $\vec{f}^{VI} \in \mathbb{R}^{257\times768}$ and $\vec{f}^{VT} \in \mathbb{R}^{77\times768}$ from fMRI voxels, which are supervised by coarse- and fine-grained features, $\vec{f}^T$ and $\vec{f}^I$, respectively. We adopt the Mean Squared Error loss for the accurate prediction of CLIP embeddings:
\begin{equation}
    % \mathcal{L}_{highlevel} = \frac{1}{B}\sum^{B}_{b=1}(||\vec{f}^{T}_{b}-\vec{f}^{VT}_{b}||^2_2+||\vec{f}^{I}_{b}-\vec{f}^{VI}_{b}||^2_2).
    \mathcal{L}_{H} = ||\vec{f}^{T}-\vec{f}^{VT}||^2_2+||\vec{f}^{I}-\vec{f}^{VI}||^2_2.
\end{equation}

% Due to the low signal-to-noise ratio of fMRI, directly learning a mapping between fMRI and the semantic condition is prone to overfitting, which is essentially caused by information asymmetry. Considering the lower feature dimensionality and robust semantic information of CLIP, and given that CLIP has been shown to align with the human visual system\cite{yang2024brain} and possesses superior explanatory power for cortical activity\cite{zhouclip}, we propose the Cognitive Integration Module and Cognitive Correlation Module to introduce additional CLIP-based fine-grained semantic information to avoid overfitting and information asymmetry.

% 同时解码多粒度high-level语义和low-level details，this pipeline achieves the balance between 语义保真度and low-level attributes such as spatial structure and color composition。
By simultaneously decoding multi-granularity high-level semantics and low-level details, this pipeline achieves a balance between semantic fidelity and low-level attributes such as spatial structure and color composition.
However, directly learning a mapping between fMRI and the semantic features is prone to overfitting~\cite{luanimate}, which is essentially caused by information asymmetry.
To tackle this issue, we propose CIM and CCM to explore richer semantic patterns inspired by human visual cognition, thereby alleviating the modality gap.
% 目的 语义拿出来说 我们和其他pipeline的区别

}

\subsection{Cognitive Integration Module}
\label{sec:semantic-decoding cim}
% kmeans可能是语义类别下的子模式
% \textbf{1)} Cognitive Integration Module (CIM) designs region aggregation and enhances feature hierarchical diversity to alleviate missing attributes or objects.\\
% 

% 我们首次引入额外语义  如何体现
% Given that the image semantic feature of CLIP primarily consists of local patch and global semantic information, 
The information asymmetry between fMRI and images partly arises from the fact that humans perceive objects regionally, guided by holistic prior knowledge~\cite{koffka1922perception}, whereas images are typically processed at the patch or global level.
% For diverse local cognitive information, we propose the Cognitive Integration Module (CIM) to increase the granularity of representations and compensate for the inadequate characterization.
% To explore more region-level semantic information, we extract local sub-pattern features using both supervised and unsupervised methods for cognitive integration as shown in Fig. \ref{fig:cim}.

\begin{figure}[tb]
  \centering
  \includegraphics[height=3.9cm]{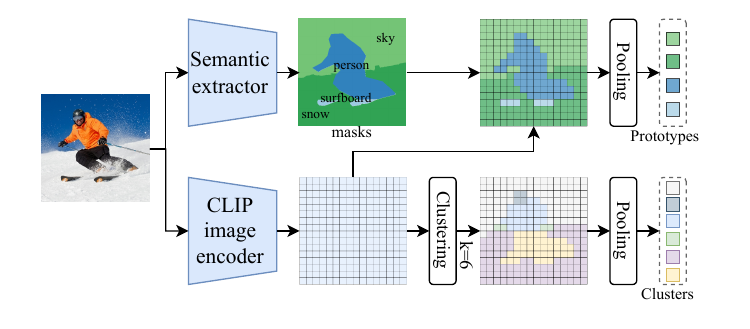}
  \caption{ Details of the CIM module. By employing panoptic segmentation masks and unsupervised clustering, we aggregate hierarchical regional semantics to simulate cognitive integration, improving intra-region semantic consistency.
  % category prototype features are generated to simulate cognitive integration.
  }
  \label{fig:cim}
\end{figure}

To mitigate this issue, we propose the Cognitive Integration Module (CIM), which is shown in Fig.~\ref{fig:cim}, to enhance regional fMRI representations and reduce information asymmetry. It simulates human prior knowledge via prototypes from annotated semantic masks, and complements finer-grained patterns from unsupervised clustering.

% For the incorporation of local semantic consistency, we first obtain the binary panoptic segmentation mask $\vec{M} \in \mathbb{R}^{H\times W}$ of the image, and subsequently extract prototype features $\vec{p}\in \mathbb{R}^{C\times768}$ with mask pooling to emulate human perception of locally holistic structures within the stimulus image. We construct the patch-level semantic mask from the panoptic segmentation mask via majority voting. This process ensures that each patch is assigned to the category that best represents its local semantic content. The category prototype for sub-pattern feature is then computed by aggregating the token features associated with the selected category, as follows:

\textbf{Prototyping.}
CIM aggregates regional representations under the prior guidance of the panoptic segmentation masks.
We convert them into patch-level masks $\vec{M} \in \mathbb{R}^{H\times W}$ ($H$ and $W$ denote patch token counts in height and width.) via majority voting, ensuring that each patch belongs to only one category.
CIM derives the prototype $p_m$ for semantic category $m$ from CLIP visual features $f^{VI}$ via mask pooling, representing aggregated regional information:
\begin{equation}
\boldsymbol{p}_{m} = \frac{{\sum{\vec{f}^{VI}_{i,j} \cdot \mathbbm{1} \left( {\boldsymbol{M}_{i,j} = m} \right)}}\,}{\sum{\mathbbm{1}\left( {\boldsymbol{M}_{i,j} = m} \right)\,}},m\in [0,C-1],
\end{equation}
where $\mathbbm{1}$ is an indicator denoting whether current patch belongs to category $m$, $C$ denotes the number of semantic categories.

% \textbf{Clustering.} To further exploit local semantic information in an unsupervised manner, we utilize the K-means clustering algorithm~\cite{kmeans} to adaptively identify meaningful features. Specifically, all token features are grouped into $k$ clusters (e.g., person, sky, helmet, snow, as illustrated in Fig.~\ref{fig:cim}). Each cluster center for sub-pattern feature, denoted as $\vec{c}_k \in \mathbb{R}^{1\times768}$, is computed by pooling all token features within the $k$-th cluster. The clustering process allows the model to capture more fine-grained regional semantics that may be overlooked by annotated labels, thereby enriching the overall representation through both supervised and unsupervised semantic sub-pattern features.

\textbf{Clustering.}
While semantic categories encode human prior knowledge, they overlook regional commonalities such as color and attributes. CIM complements them by adaptively aggregating regions through unsupervised clustering to capture finer-grained semantics.
Specifically, CIM applies K-means~\cite{kmeans} clustering to $f^{VI}$ into $K$ clusters, with the cluster center denoted as $\vec{c}_k \in \mathbb{R}^{1\times768}$ for cluster $k$.

% Then, we force the sub-pattern features of category prototype and clustering centers as additional supervision objectives during fMRI feature learning:
Both prototypes and clusters fundamentally capture multi-level visual region patterns. CIM employs MSE loss to incorporate them as additional supervision, enhancing semantic granularity for fMRI decoding.
Furthermore, we introduce contrastive learning to guide fMRI features in distinguishing different region semantics, thereby enforcing intra-region semantic consistency.
% with contrastive learning enforcing intra-region semantic consistency.
% This alleviates information asymmetry and improves fMRI representations.
% Additionally, contrastive learning is introduced to guide the fMRI features in distinguishing semantic differences at varying granularities, thereby improving the representation of fMRI.
We denote the training process as $\mathcal{L}_{CIM}$:
% \begin{equation}
% \begin{align}
% \label{tau}
% \mathcal{L}_{CIM} = &\sum^{C}_{m=1}\frac{1}{N_c}\sum_{j\in \mathcal{C}_m}(||\vec{f}^{VI}_j-\vec{p}_m||^2_2- \frac{exp(\frac{\vec{f}^{VI}_j\cdot\vec{p}_m}{\tau})}{\sum_i^{C}exp(\frac{\vec{f}^{VI}_j\cdot\vec{p}_i}{\tau})}) + \notag \\
% &\sum^{K}_{k=1}\frac{1}{N_k}\sum_{j\in \mathcal{C}_k}(||\vec{f}^{VI}_j-\vec{c}_k||^2_2 - \frac{exp(\frac{\vec{f}^{VI}_j\cdot\vec{c}_k}{\tau})}{\sum_i^{K}exp(\frac{\vec{f}^{VI}_j\cdot\vec{c}_i}{\tau})}),
% \end{align}
\begin{align}
\label{tau}
\mathcal{L}_{CIM} = &\sum^{C}_{m=1}\frac{1}{N_c}\sum_{j\in \mathcal{C}_m}(||\vec{f}^{VI}_j-\vec{p}_m||^2_2- \frac{exp(\frac{\vec{f}^{VI}_j\cdot\vec{p}_m}{\tau})}{\sum_{i=1}^{C}exp(\frac{\vec{f}^{VI}_j\cdot\vec{p}_i}{\tau})})  \notag \\
&+ \sum^{K}_{k=1}\frac{1}{N_k}\sum_{j\in \mathcal{C}_k}(||\vec{f}^{VI}_j-\vec{c}_k||^2_2 - \frac{exp(\frac{\vec{f}^{VI}_j\cdot\vec{c}_k}{\tau})}{\sum_{i=1}^{K}exp(\frac{\vec{f}^{VI}_j\cdot\vec{c}_i}{\tau})}),
\end{align}
% \end{equation}
where $\mathcal{C}_m$ and $\mathcal{C}_k$ are the sets of tokens belonging to the corresponding prototype $m$ and cluster $k$, respectively.
$N_c$ and $N_k$ denote the number of their respective token sets.
$\tau$ is the temperature parameter in contrastive learning.
% $N_c$ and $N_k$ are the numbers of tokens in the $c$-th category and $k$-th cluster, respectively, and $\mathcal{C}_m$ and $\mathcal{C}_k$ are the sets of tokens belonging to the corresponding category and cluster.

By integrating multi-level semantic aggregation with contrastive learning, CIM enhances semantic consistency within regions, leading to faithful reconstruction of object details.

% {\color{rev}
% CIM simulates the integration and segmentation of complex objects and regions by the inferior temporal cortex, thereby enhancing the coherence of attributes or objects within each region and mitigating the omission of objects or attributes in the generated reconstructions.
% By semantic sub-pattern mining, CIM introduces supplementary semantic information to fMRI representations in both supervised and unsupervised ways, alleviating inadequate characterization due to modality information asymmetry.
% }

\subsection{Cognitive Correlation Module}
\label{sec:semantic-decoding ccm}
% \textbf{2)} Cognitive Correlation Module (CCM) introduces topological correlation to alleviate confusion caused by spatial layout and object relationships.\\
% 结合脑区和相关性
% CIM alleviates object or attribute omission, yet it is also crucial to maintain the layout between objects consistent with the stimuli for adequate characterization. Previous methods~\cite{han2024mindformer,takagi2023high,wang2024mindbridge} neglect this aspect, leading to spatial layout inconsistencies and structural misplacement. Hence, CCM incorporates topological relationships to steer fMRI representation learning, with the goal of acquiring a more comprehensive context and consistent relationships among objects.

CIM introduces multi-level semantics to ensure intra-region consistency, yet it is also crucial to capture contextual relationships for fMRI decoding.
To this end, we propose the Cognitive Correlation Module (CCM), enabling fMRI features to learn the topological relationships among semantics motivated by the human differential perception process~\cite{masland2012neuronal,poggio1990network}.

Specifically, CCM learns fine- and coarse-grained semantic contextual relationships from CLIP visual and text features, respectively.
The relationships are established via Cosine distance, denoted as $\epsilon(\cdot)$.
The training objective of CCM is:
\begin{equation}
    \mathcal{L}_{CCM} = ||\epsilon(\vec{f}^I) - \epsilon(\vec{f}^{VI})||_1+||\epsilon(\vec{f}^T) - \epsilon(\vec{f}^{VT})||_1,
\end{equation}
where the fine-grained relationship matrix of $\vec{f}^{VI}$ is $\mathbb{R}^{257\times257}$ and coarse-grained relationship matrix of $\vec{f}^{VT}$ is $\mathbb{R}^{77\times77}$.

% Specifically, the topological relationship operator is defined as $\epsilon(\cdot)$, here we use the Cosine similarity, and the relationship matrix of CLIP features is utilized to guide those of fMRI features, thereby enriching the contextual representation of fMRI:
% \begin{equation}
%     \mathcal{L}_{CCM} = ||\epsilon(\vec{f}^I) - \epsilon(\vec{f}^{VI})||_1+||\epsilon(\vec{f}^T) - \epsilon(\vec{f}^{VT})||_1,
% \end{equation}
% where the relationship matrix is $\mathbb{R}^{257\times257}$ for fine-grained and $\mathbb{R}^{77\times77}$ for coarse-grained features.

The topological relationships of visual stimuli in the CLIP representation space serves as prior knowledge to guide fMRI decoding, mitigating the spatial layout inconsistencies and structural misplacement.
CCM further facilitates the fidelity of inter-region relationships between elements in visual stimuli based on more complete elements or attributes from CIM.

% Finally, BrainCognizer is applicable to both specific-subject and cross-subject scenarios. When facing the individual variability~\cite{clipmused} across subjects, our model can also mitigate the inadequate characterization of fMRI, improving visual reconstruction quality and enhancing the generalization ability across subjects.

\begin{table*}[t]\small
\centering
\caption{Quantitative comparison of reconstruction performance with SOTA models. {\color{rev}Some methods are excluded because they are not open-source or used additional data.} All metrics are averaged over subjects 1, 2, 5, and 7. The best results are highlighted in bold, and the second-best results are underlined. Arrows indicate that higher or lower values are preferable.}
\resizebox{180mm}{!}{
\begin{tabular}{ ccccccccc }
\toprule
\multirow{2}{*}{Method}&\multicolumn{4}{c}{Low-Level}&\multicolumn{4}{c}{High-Level}\\
\cmidrule(lr){2-5} \cmidrule(lr){6-9}
& PixCorr $\uparrow$ & SSIM $\uparrow$ & Alex(2) $\uparrow$ & Alex(5) $\uparrow$ & Incep $\uparrow$ & CLIP $\uparrow$ & EffNet-B $\downarrow$ & SwAV $\downarrow$ \\
\midrule

% \multicolumn{9}{l}{Brain decoding average results across the same 4 participants with four models.}\\
% \multicolumn{9}{c}{Specific-subject setting}\\
% \hline
Ozcelik et al. \cite{ozcelik2022reconstruction} (IJCNN 2022) & .126 & .135 & 68.9\% & 81.2\% & 76.4\% & 76.3\% & .879 & .579 \\
Mind-Reader \cite{lin2022mind} (NIPS 2022) & .104 & .294 & 70.9\% & 83.9\% & 78.2\% & 78.1\% & .853 & .463 \\
% Takagi et al.~\cite{takagi2023high} (CVPR 2023)& - & - & 83.0\% & 83.0\% & 76.0\% & 77.0\% & - & - \\
Takagi et al.~\cite{takagi2023high} (CVPR 2023) & .222 & .318 & 83.0\% & 83.0\% & 76.0\% & 77.0\% & .916 & .578 \\
% Gu et al.~\cite{gu2024decoding} (MIDL 2023)&.150&.325 &-&-&-&-&.862& .465\\
Gu et al.~\cite{gu2024decoding} (MIDL 2023)&.082 & .297 & 68.9\% & 79.9\% & 75.2\% & 70.4\% & .901 & .501 \\
Brain-Diffuser~\cite{ozcelik2023natural} (Sci. Rep 2023)& .254 & \textbf{.356} & 94.2\% & 96.2\% & 87.2\% & 91.5\% & .775 & .423 \\
MindEye~\cite{scotti2024reconstructing} (NeurIPS 2023) & \underline{.309} & .323 & \textbf{94.7\%} & \textbf{97.8\%} & \underline{93.8\%} & 94.1\% & \textbf{.645} & \underline{.367} \\
% Brain-Streams\cite{joo2024brainstreamsfMRItoimagereconstructionmultimodal} & .342 & .365 & 94.7\% & 97.0\% & 94.0\% & 95.2\% & .651 & .357 \\
MindDiffuser \cite{lu2023minddiffuser} (MM 2023)& .256 & .344 & 85.2\% & 84.3\% & 78.4\% & 79.1\% & .884 & .551 \\
MindBridge ~\cite{wang2024mindbridge} (CVPR 2024)& .148 & .259 & 86.9\% & 95.3\% & 92.2\% & \underline{94.3\%} & \underline{.713} & .413 \\
DREAM \cite{xia2024dream} (WACV 2024) & .274 & .328 & 93.9\% & 96.7\% & 93.4\% & 94.1\% & \textbf{.645} & .410 \\
% MindFormer ~\cite{han2024mindformer} (arxiv 2024) & .241 & .352 & 93.5\% & 97.5\% & 93.5\% & 93.6\% & .659 & .356 \\
\textbf{BrainCognizer}  & \textbf{.319} & \underline{.346} & \underline{94.5\%}& \underline{97.5\%} & \textbf{94.8\%} & \textbf{95.3\%} & \textbf{.645} & \textbf{.365} \\

\bottomrule
\end{tabular}}
\label{mainresult}
\end{table*}

\subsection{Training And Inference}
\label{sec:inference}

During the training process, the above two modules simultaneously provide semantic guidance for fMRI decoding. 
% Our BrainCognizer integrates losses of two modules in each sample for optimization, besides adhering to the training configurations (low-level loss $\mathcal{L}_{lowlevel}$ and high-level loss $\mathcal{L}_{highlevel}$) of baselines:
In summary, the final training objective of BrainCognizer is:
\begin{equation}
    \mathcal{L}_{total} = \mathcal{L}_{L} + \mathcal{L}_{H} + \mathcal{L}_{CIM} + \mathcal{L}_{CCM}.
\end{equation}

During inference, only brain signals are needed and processed by the encoder to obtain the low-level feature $\vec{f}^{VL}$ and high-level semantic features $\vec{f}^{VT}$, $\vec{f}^{VI}$.
The low-level feature is subsequently input to the VAE decoder, producing a low-level image with blurred color and contour information, which serves as the initialization for Versatile Diffusion. Finally, the semantic features serve as conditions to provide controllable guidance for image generation in the diffusion model.

\section{Experiments}
\subsection{Experimental Settings}
\subsubsection{Dataset}
% st：follow by other works
% st：脑区
{\color{rev}
We conduct experiments on the Natural Scenes Dataset (NSD)~\cite{allen2022massive}, the largest public fMRI dataset for image reconstruction.
% Following previous methods~\cite{gu2024decoding,han2024mindformer,ozcelik2023natural,scotti2024reconstructing,takagi2023high,wang2024mindbridge}, we utilize brain fMRI signals collected from four human participants who viewed three repetitions of complex natural images from the MS-COCO dataset~\cite{coco}. 
Following previous methods~\cite{gu2024decoding,han2024mindformer,ozcelik2023natural,scotti2024reconstructing,takagi2023high,wang2024mindbridge}, we use fMRI data from four participants who viewed three repetitions of complex natural images from the MS-COCO dataset~\cite{coco}.
% As our study focuses on the visual cortex, we adopt the official \texttt{nsdgeneral} region of interest (ROI) mask, which spans early to high-level visual areas, consistent with prior works.
We adopt the official \texttt{nsdgeneral} region of interest (ROI) mask for the visual cortex, which spans early to high-order visual areas, consistent with prior works.
The fMRI data within the selected ROI are flattened into a 1D voxel sequence for decoding.
For each subject, the training set consists of 8,859 unique image stimuli and 24,980 fMRI trials (with up to three per image). 
% In the test set, 982 images and 2,770 fMRI trials are shared across all four subjects.
The test set contains 982 shared images and 2,770 trials across all four subjects.
}

\subsubsection{Evaluation Metrics}
% st：三个方面，相似性、分类、距离
{\color{rev}
% We use four metrics to evaluate low-level reconstruction quality, including Pixelwise Correlation (PixCorr) and Structural Similarity Index Metric (SSIM)\cite{wang2004image}, and two-way identification based on features from the second and fifth layers of AlexNet\cite{krizhevsky2012imagenet}.
We evaluate low-level reconstruction quality using four metrics: Pixelwise Correlation (PixCorr), Structural Similarity Index (SSIM)~\cite{wang2004image}, and two-way identification based on features from AlexNet’s second and fifth layers~\cite{krizhevsky2012imagenet}.
For high-level reconstruction evaluation, we adopt four additional metrics: two-way identification with Inception~\cite{szegedy2016rethinking} and CLIP~\cite{radford2021learning}, as well as average correlation distance with EfficientNet-B1~\cite{tan2019efficientnet} and SwAV-ResNet50~\cite{caron2020unsupervised}, following previous studies~\cite{gu2024decoding,han2024mindformer,ozcelik2023natural,scotti2024reconstructing,takagi2023high,wang2024mindbridge}.
}

\subsubsection{Implementation Details}
The experiments are conducted using 2 NVIDIA RTX 4090 GPUs, with a batch size set to 50 per GPU. Our learning rate is set to 3e-4, $\tau$ in Eq.~\ref{tau} is 5e-2, and we use the AdamW strategy to train for 200 epochs.
% BrainCognizer sets the hyperparameter $K^I$ and $K^T$ to 15 and 5 in CIM.

\begin{figure}[tb]
  \centering
  \includegraphics[height=8.5cm]{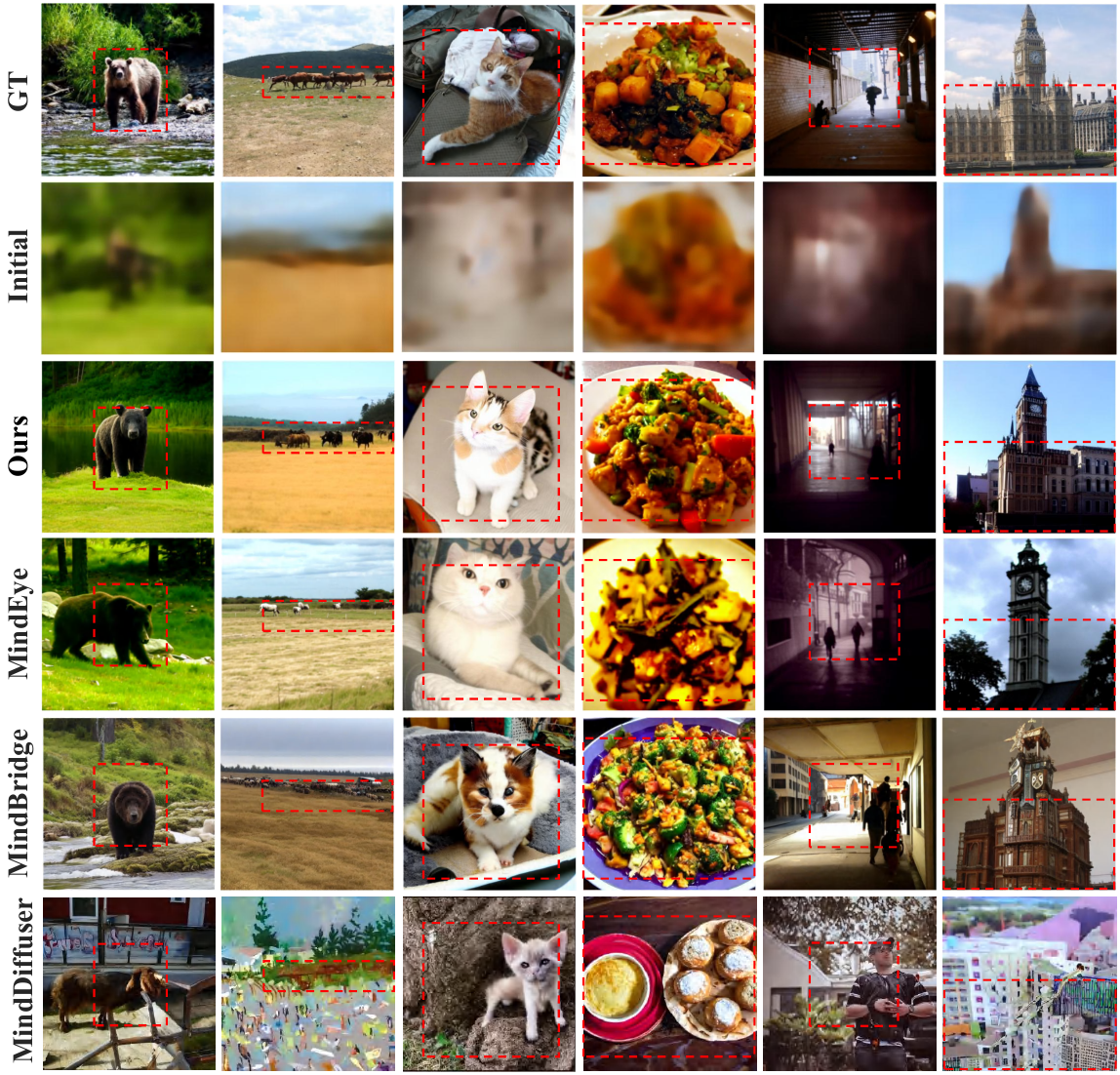}
  \caption{Qualitative comparison of reconstruction examples on subject 1 given input fMRI signals.
  The highlighted regions indicate comparison.
  We reconstruct more fine-grained semantic attributes (e.g., the 1st and 6th columns), while also preserve more consistent spatial layout (e.g., the 5th column).
  }
  \label{fig:reconstruction}
\end{figure}

\subsection{Comparisons With SOTA Methods}
\subsubsection{Quantitative Comparison}
We report the average performance of the model across four subjects in terms of both high-level semantic relevance and low-level pixel similarity. 
% As summarized in Tab.~\ref{mainresult}, BrainCognizer outperforms state-of-the-art methods across nearly all evaluation metrics. Overall, BrainCognizer demonstrates more robust performance, achieving state-of-the-art (SOTA) results on five metrics and ranking second on the remaining three.
As shown in Tab.~\ref{mainresult}, BrainCognizer exhibits robust performance, achieving state-of-the-art (SOTA) results on five evaluation metrics and ranking second on the remaining three.
% It exhibits a significant advantage in high-level semantic evaluations, further demonstrating the effectiveness of our proposed strategy for incorporating semantic information. 
It exhibits a significant advantage in high-level semantic evaluations, indicating that incorporating multi-level intra-region semantics and inter-region semantic correlations enables more comprehensive decoding of visual semantics from fMRI.

Notably, under the same CLIP and diffusion model settings as MindBridge, BrainCognizer achieves an average improvement of 6.4\% across the eight evaluation metrics. Compared with MindEye, we do not introduce any additional network structures, yet still achieve even superior performance without incurring extra computational cost.

BrainCognizer simulates the human visual cognition to effectively extract meaningful semantics from fMRI signals and achieves the highest CLIP score, thereby providing a powerful learning paradigm for fMRI representation that bridges the modality gap through alignment with CLIP embeddings.
\subsubsection{Qualitative Comparison}
% 图5中重建示例佐证了我们方法的有效性。我们的方法区域内语义一致性更好，能够保留更多的图片细节属性,例如第一列中的熊的耳朵及第六列的building.此外,我们的方法也保留了更多的区域间语义相关性,能够保留更一致的空间和上下文布局,例如第四列蔬菜的排列和第五列空间布局.
The reconstruction images in Fig.~\ref{fig:headrec} and Fig.~\ref{fig:reconstruction} demonstrate the effectiveness of BrainCognizer.
BrainCognizer preserves intra-region semantic consistency and image details, such as the bear's ears (1st column) and the building (6th column). 
It also better preserves inter-region semantic correlations  reflected in consistent spatial and contextual layouts, such as the vegetable arrangement (4th column) and the lighting structure (5th column). Moreover, BrainCognizer uses fMRI-decoded images (2nd row) to initialize diffusion, reducing randomness and enhancing consistency.

\subsection{Robustness across subjects}
Inter-subject variability in brain anatomy and neural responses leads to differences in fMRI signals, even under identical visual stimuli. To evaluate BrainCognizer's generalizability, we reconstruct test images from subjects 1, 2, 5, and 7.
As shown in Tab.~\ref{subject_table}, despite the completely different training sets across the four subjects, we achieve comparable performance, especially on high-level metrics.
% Results in Fig.~\ref{fig:1257} show that, despite minor deviations caused by individual differences, we achieves consistent semantic and structural alignment across subjects.
The examples in Fig.~\ref{fig:1257} further support this finding, showing only minor stylistic differences across individuals, which indicates that our method generalizes well across subjects.

\begin{table}[t]
\centering
\caption{Ablation study of our components. The best result is highlighted in bold, and the second-best is underlined.}
\resizebox{85mm}{!}{
\begin{tabular}{llcccc}
\toprule
\multirow{2}{*}{Modules} 
    & CIM & \ding{55} & \ding{51} & \ding{55} & \ding{51} \\
    & CCM & \ding{55} & \ding{55} & \ding{51} & \ding{51} \\
\midrule
\multirow{4}{*}{Low-Level} 
    & PixCorr $\uparrow$ & .394 & .398 & \underline{.405} & \textbf{.406} \\
    & SSIM $\uparrow$ & .347 & .351 & \textbf{.367} & \underline{.364} \\
    & Alex(2) $\uparrow$ & 96.9\% & \textbf{97.7\%} & 96.9\% & \underline{97.5\%} \\
    & Alex(5) $\uparrow$ & 98.1\% & \underline{98.5\%} & 97.8\% & \textbf{98.6\%} \\
\midrule
\multirow{4}{*}{High-Level} 
    & Incep $\uparrow$ & 92.8\% & \underline{94.9\%} & 93.8\% & \textbf{95.3\%} \\
    & CLIP $\uparrow$ & 93.9\% & \underline{95.1\%} & 94.3\% & \textbf{96.0\%} \\
    & EffNet-B $\downarrow$ & .715 & \underline{.649} & .657 & \textbf{.641} \\
    & SwAV $\downarrow$ & .407 & \underline{.371} & .379 & \textbf{.360} \\
\bottomrule
\end{tabular}
}
\label{ablationtable}
\end{table}

\begin{figure}[tb]
  \centering
  \includegraphics[height=4.9cm]{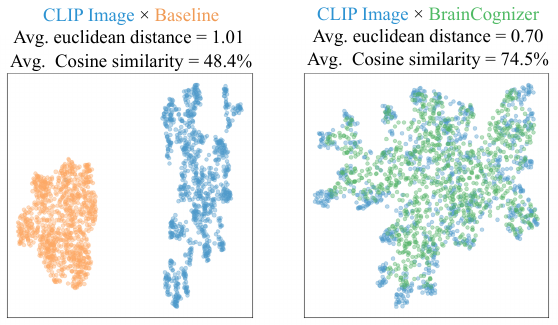}
  \caption{UMAP~\cite{mcinnes2018umap} visualizations of the CLIP (blue), Baseline (orange), and our BrainCognizer (green).
  The UMAPs are estimated from 982 samples in the testing set.
  }
  \label{fig:duiqi}
\end{figure}

\subsection{Ablation Study}
% \subsubsection{Module contribution}
% \begin{table}[t]
% \centering
% \caption{Ablation study of our components. Bold font denotes the best result.}
% \resizebox{90mm}{!}{
% \begin{tabular}{ cccccccccc }
% \toprule
% \multicolumn{2}{c}{Modules}&\multicolumn{4}{c}{Low-Level}&\multicolumn{4}{c}{High-Level}\\
% \cmidrule(lr){1-2} \cmidrule(lr){3-6} \cmidrule(lr){7-10}
% CIM & CCM & PixCorr $\uparrow$ & SSIM $\uparrow$ & Alex(2) $\uparrow$ & Alex(5) $\uparrow$ & Incep $\uparrow$ & CLIP $\uparrow$ & EffNet-B $\downarrow$ & SwAV $\downarrow$ \\
% \midrule

% & & .394 & .347 & 96.9\% & 98.1\% & 92.8\% & 93.9\% & .715 & .407 \\
% $\checkmark$ & &.398 & .351 & 97.7\% & 98.5\% & 94.9\% & 95.1\% & .649 & .371 \\
%  &$\checkmark$ &.405 & .367 & 96.9\% & 97.8\% & 93.8\% & 94.3\% & .657 & .379 \\
% $\checkmark$&$\checkmark$ & .403 & .362 & 97.5\% & 98.6\% & 95.0\% & 95.8\% & .648 & .362 \\
% \bottomrule
% \end{tabular}}
% \label{ablationtable}
% \end{table}
{\color{rev}
We progressively incorporate each module into BrainCognizer to evaluate their individual contributions to the final performance on Subject~1, as shown in Tab.~\ref{ablationtable}. Our proposed pipeline serves as a strong baseline, demonstrating balanced performance across all eight evaluation metrics.
% Overall, the complete method achieves an average improvement of 2.6\% across all metrics. Specifically, CIM and CCM contribute average improvements of 1.9\% and 1.6\%, respectively, over the baseline. 

Since the human visual system is insensitive to low-level variations, high-level metrics are considered more indicative of perceptual reconstruction quality~\cite{neuropictor}.
% Notably, our method enhances reconstruction performance primarily by extracting additional intra-region semantic consistency and inter-region semantic correlation from fMRI signals, which directly benefits high-level metrics.
On high-level metrics, our method achieves an average improvement of 4.2\%, with CIM and CCM individually contributing 3.4\% and 2.5\%, respectively, while maintaining performance on low-level metrics. The integration of the proposed modules enables more effective preservation of rich semantic information from visual stimuli and improves the alignment between fMRI representations and reconstructed images. Ablation studies demonstrate that incorporating additional semantic and relational information is crucial for accurately aligning reconstructed images with the original visual stimuli at a fine-grained level.

}

% {\color{red}CIM and CCM facilitate high-level information learning. CIM improves object or attribute omission, while CCM alleviates spatial layout inconsistencies and relational misalignment.}

{\color{rev}
% To verify the effect of our semantic constraints on cross-modal feature alignment, we visualize the semantic space of CLS tokens for all test samples in Fig.~\ref{fig:duiqi}, which demonstrates that our method captures the semantic encoding capabilities of CLIP more effectively than the baseline.
% To verify the effect of our semantic constraints on cross-modal feature alignment, we visualize the semantic space of CLS tokens in Fig.~\ref{fig:duiqi}, which demonstrates that our method captures the semantic encoding capabilities of CLIP more effectively than the baseline.
% As shown in Fig.~\ref{fig:duiqi}, the visualization of the CLS token space indicates that our semantic constraints improve cross-modal alignment and more effectively leverage CLIP’s semantic encoding.
As shown in Fig.~\ref{fig:duiqi}, our semantic constraints improve cross-modal alignment and better leverage CLIP’s semantic encoding.
Furthermore, we calculate the Cosine similarity between fMRI-derived semantic features and image semantic features, achieving a 53.8\% improvement over the baseline. The Euclidean distance also decreases by 30.4\%, indicating that BrainCognizer successfully aligns the two embedding spaces and reduces the modality gap.
Leveraging brain-inspired perception, our method fuses fine-grained semantics, enhancing generalization and mitigating overfitting.
% Leveraging brain-inspired perception, our method enables the effective fusion of fine-grained semantic information, leading to improved generalization and reduced overfitting.}

\begin{table}[t]
\centering
\caption{Results on subjects 1, 2, 5, and 7.}
\resizebox{85mm}{!}{
\begin{tabular}{llcccc}
\toprule
\multicolumn{2}{c}{Metric} & Sub 1 & Sub 2 & Sub 5 & Sub 7 \\
\midrule
\multirow{4}{*}{Low-Level}
    & PixCorr $\uparrow$ & .406 & .327 & .277 & .264 \\
    & SSIM $\uparrow$ & .364 & .348 & .338 & .335 \\
    & Alex(2) $\uparrow$ & 97.5\% & 95.7\% & 92.8\% & 92.0\% \\
    & Alex(5) $\uparrow$ & 98.6\% & 97.9\% & 97.4\% & 95.9\% \\
\midrule
\multirow{4}{*}{High-Level}
    & Incep $\uparrow$ & 95.3\% & 94.8\% & 95.6\% & 93.3\% \\
    & CLIP $\uparrow$ & 96.0\% & 94.6\% & 96.7\% & 93.9\% \\
    & EffNet-B $\downarrow$ & .641 & .650 & .619 & .669 \\
    & SwAV $\downarrow$ & .360 & .364 & .353 & .381 \\
\bottomrule
\end{tabular}
}
\label{subject_table}
\end{table}

\begin{figure}[tb]
  \centering
  \includegraphics[height=8.2cm]{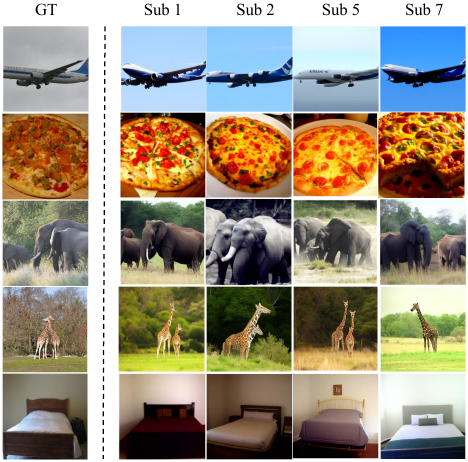}
  \caption{Our reconstruction visualization on subjects 1, 2, 5, and 7. These results demonstrate that Braincognizer exhibits inter-subject consistency and generalization.}
  \label{fig:1257}
\end{figure}

% \begin{figure*}[tb]
%   \centering
%   \includegraphics[height=2.6cm]{heatmap.pdf}
%   \caption{The sensitivity of $K$ ($K^T$ for text, $K^I$ for image) on performance improvements of high-level metrics (\%) on Subject 1. 
%   }
%   \label{fig:k_heatmap}
% \end{figure*}

% \subsubsection{Hyperparameter analysis.}
% For the hyperparameter $K$ of K-means in CIM, we conduct an analysis to evaluate its sensitivity on high-level metrics since that it mainly affects high-level metrics.
% As shown in Fig.~\ref{fig:k_heatmap}, we evaluate \( K^{I} = \{10, 15, 20\} \) (total 257 tokens) and \( K^{T} = \{3, 5, 10\} \) (total 77 tokens), and demonstrate that the clustering in CIM is robust and effective at the high level metrics. The final model employs \( K^{T} = 5 \) and \( K^{I} = 15 \) for the best performance.

\subsection{Hyperparameter analysis}
% To investigate the sensitivity of the number of clusters, Tab.~\ref{tab:braincognizer_k} presents a comparative evaluation of BrainCognizer under different numbers of clusters ($k = 5, 10, 15, 20, 25$). The method shows low sensitivity to the choice of $k$, and maintains consistent low-level pixel and structural similarity. 
To investigate the sensitivity of cluster numbers in CIM, Tab.~\ref{tab:braincognizer_k} presents a comparative evaluation of BrainCognizer under different cluster numbers, which shows low sensitivity of $k$ with stable low-level pixel and structural similarity. 
The K-means clustering provides finer-grained semantics compared to prototypes, enabling the integration of hierarchical semantic cues across different values of $k$. Cluster sizes of $k = 15$, $20$, and $25$ achieve a favorable trade-off between diversity and representational capacity. Based on this analysis, $k = 15$ is adopted in the final model.

\begin{table}[t]
\centering
\caption{
Performance of BrainCognizer with different numbers of clusters \(k\) on Subject 1. Bold values highlight the best results for each metric.
}
\resizebox{85mm}{!}{
\begin{tabular}{l|c|ccccc}
\toprule
\multirow{2}{*}{Metric} 
& \multirow{2}{*}{Baseline} 
& \multicolumn{5}{c}{BrainCognizer} \\
\cmidrule(lr){3-7}
& & k=5 & k=10 & k=15 & k=20 & k=25 \\
\midrule
PixCorr $\uparrow$       & 0.394 & 0.405 & 0.405 & \textbf{0.406} & 0.405 & \textbf{0.406} \\
SSIM $\uparrow$          & 0.347 & 0.362 & 0.361 & \textbf{0.364} & 0.363 & \textbf{0.364} \\
Alex(2) $\uparrow$       & 96.90\% & 97.6\% & \textbf{97.7\%} & 97.5\% & 97.6\% & 97.5\% \\
Alex(5) $\uparrow$       & 98.10\% & \textbf{98.6\%} & 98.5\% & \textbf{98.6\%} & \textbf{98.6\%} & 98.5\% \\
Incep $\uparrow$         & 92.80\% & 95.0\% & 94.9\% & 95.3\% & 95.1\% & \textbf{95.5\%} \\
CLIP $\uparrow$          & 93.90\% & 95.5\% & 95.6\% & \textbf{96.0\%} & 95.9\% & 95.9\% \\
EffNet-B $\downarrow$    & 0.715 & 0.644 & 0.641 & 0.641 & \textbf{0.637} & \textbf{0.637} \\
SwAV $\downarrow$        & 0.407 & 0.362 & 0.361 & 0.360 & \textbf{0.359} & 0.360 \\
\bottomrule
\end{tabular}}

\label{tab:braincognizer_k}
\end{table}

% \begin{figure*}[tb]
%   \centering
%   \includegraphics[height=12.7cm]{figure/brainvisual2.pdf}
%   \caption{ROI-wise importance in decoding information for subject 1. The weights learned by the linear regression model are projected onto the cortical surface and visualized on a flattened 2D map with ROI annotations. Redder areas indicate stronger contributions to the decoding process. Subfigures show the fitted weights for: (a) detailed information of $\vec{f}^{L}$, (b) fine-grained semantic information of $\vec{f}^{I}$, (c) coarse-grained semantic information of $\vec{f}^{T}$, (d) semantic information of prototypes, (e) correlation information of $\vec{f}^{I}$, and (f) correlation information of $\vec{f}^{T}$.}
  
%   \label{fig:visualbrain}
% \end{figure*}

\begin{figure}[tb]
  \centering
  \includegraphics[height=8.6cm]{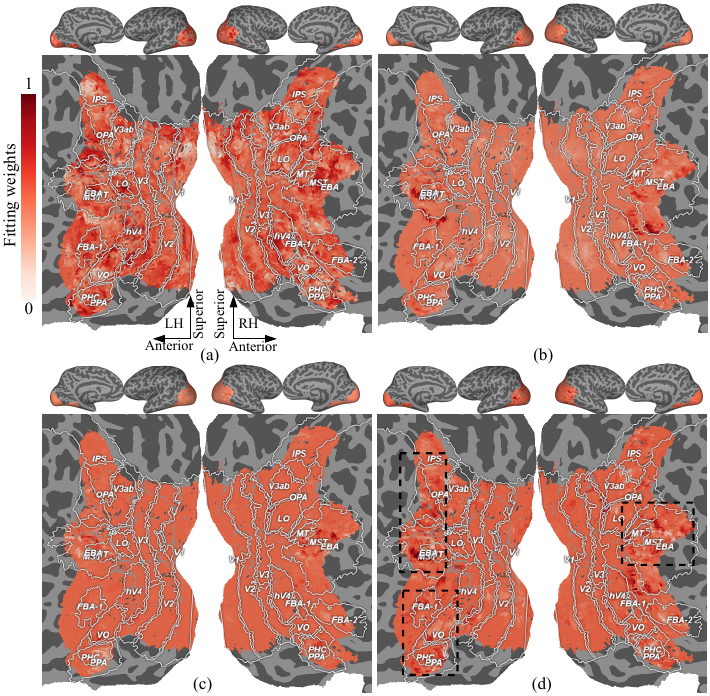}
  \caption{ROI-wise importance in decoding information.
  % The weights learned by the linear regression model are projected onto the cortical surface and visualized on a flattened 2D map with ROI annotations. 
  % Redder areas indicate stronger contributions to the decoding process. 
  Regression weights are mapped onto the 2D flat cortex map with ROI labels. Red indicates stronger decoding contributions. We visualize the activations of:
  (a) low-level features of VAE, (b) fine-grained semantic features from CLIP image encoder, (c) coarse-grained semantic features from CLIP text encoder, (d) hierarchical intra-region semantic features of CIM.}
  
  \label{fig:visualbrain}
\end{figure}

\begin{figure}[tb]
  \centering
  \includegraphics[height=4.7cm]{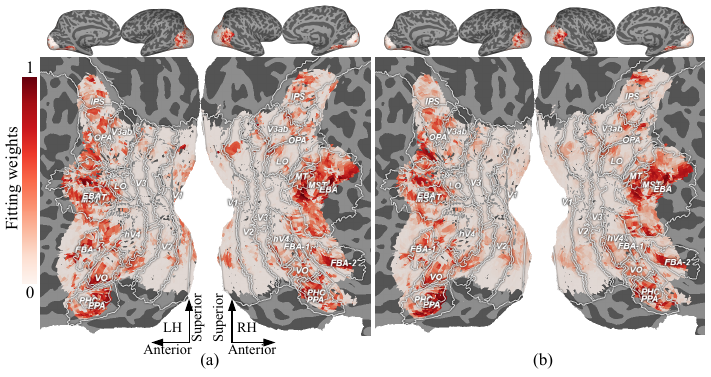}
  \caption{ROI-wise importance in decoding. 
  % The weights learned by the linear regression model are projected onto the cortical surface and visualized on a flattened 2D map with ROI annotations. 
  % Redder areas indicate stronger contributions to the decoding process. 
  We visualize the activations of: (a) correlation of coarse-grained features in CCM, (b) correlation of fine-grained features in CCM.}
  
  \label{fig:visualbrainb}
\end{figure}
{\color{rev}
\subsection{Neuroscience interpretability}
% To investigate the interpretability of BrainCognizer in neuroscience and to explore the relationship between fMRI signals and stimulus image information, we design the following experiments.
We investigate the interpretability of decoded features from a neuroscience perspective by mapping them onto brain voxel activations.
% During the feature decoding process, we employ an L2-regularized linear regression model to automatically select voxels for fitting six types of information: detailed features $\vec{f}^{L}$, fine-grained semantic features $\vec{f}^{I}$, coarse-grained semantic features $\vec{f}^{T}$, prototype features, as well as the correlation information of fine-grained semantic features $\vec{f}^{I}$ and of coarse-grained semantic features $\vec{f}^{T}$.
% We employ an L2-regularized linear regression model to automatically select voxels for fitting six types of information: detailed features $\vec{f}^{L}$, fine-grained semantic features $\vec{f}^{I}$, coarse-grained semantic features $\vec{f}^{T}$, sub-pattern semantic features, and the correlation information of fine-grained semantic features $\vec{f}^{I}$ and of coarse-grained semantic features $\vec{f}^{T}$.
We employ the L2-regularized linear regression to automatically select voxels for fitting decoded features, and project the weights onto the 3D brain cortex.
% BrainCognizer同时从low level和多粒度的high-level角度解码fMRI信息。我们将他们分别映射到大脑视觉相关皮层，从激活差异分析他们的解码(模式)。
% BrainCognizer decodes low-level and multi-granularity high-level information from fMRI, mapping them onto visual cortical areas to analyze decoding patterns based on activation differences.
BrainCognizer decodes multi-level semantic information from fMRI and maps it to visual cortical areas to analyze activation-based patterns.

% low level detailed feature  in Fig.~\ref{fig:visualbrain} (a)的激活了大部分视觉相关的皮层，尤其是早期视觉皮层，如v1到v4，主要是为fMRI解码提供了早期视觉信息 例如color, local shape, and texture information，，如图所示，low level为重建提供颜色，布局等初始化，与对应皮层相符。
% The low-level features in Fig.~\ref{fig:visualbrain}(a) activate most of the visual-related cortical areas, especially early visual areas such as V1 to V4, which are associated with color and shape.
As shown in Fig.~\ref{fig:visualbrain}(a), low-level features activate most visual cortical areas, especially early regions (V1–V4) related to color and shape.
% As shown in the second row of Fig.~\ref{fig:reconstruction}, low-level features provide initialization cues (e.g., color and layout) aligned with the functions of corresponding cortical regions.
In Fig.~\ref{fig:reconstruction} (second row), low-level features provide initialization cues (e.g., color, layout) consistent with corresponding cortical functions.
% clip的图像与文本（b和c）高级特征的激活结果相对平均，在eba等脑区上有少量高亮，体现了其和什么什么相关。
% 尽管高级特征激活视觉相关皮层，但是明显能看出其尚未挖掘出高级脑区中体素蕴含的知识。
% The ROI-wise importance in decoding CLIP image and caption features (Fig.~\ref{fig:visualbrain}(b) and (c)) shows relatively uniform activation across visual-related brain regions, with slight highlights in higher-order areas such as EBA, FBA-1, and FBA-2, indicating their involvement in processing high-level visual semantics.
As shown in Fig.~\ref{fig:visualbrain}(b) and (c), decoding CLIP image and caption features yields relatively uniform activation across visual regions, with slight emphasis in higher-order areas like EBA, FBA-1, and FBA-2, suggesting their role in visual semantic processing.
Although these features activate visual-related cortical regions, they have not fully captured the visual pattern embedded in voxels within higher-order brain areas.
% 而braincognizer能够有效缓解这一问题。我们在d中可视化cim特征对应的皮层激活结果。可以看出cim特征激活了更多高级脑区，例如eba，opa下方等，对应什么什么相关。cim为原始pipeline补充了更多什么什么信息。
BrainCognizer effectively addresses this limitation.
% In Fig.~\ref{fig:visualbrain}(d), we visualize the cortical activation of intra-region semantic features from CIM. Notably, the results reveal enhanced activation in several high-order ROIs, including the EBA, OPA, PPA, and IPS, which are known to be involved in advanced visual processing and semantic integration.
In Fig.~\ref{fig:visualbrain}(d),  the intra-region semantic features from CIM elicit stronger activation in several high-order ROIs, including the EBA, OPA, PPA, and IPS, which are involved in advanced visual processing and semantic integration.

% 在cim基础之上，图9中展示的ccm补充了更多空间布局与上下文关系。例如，相比图8的bc，ccm在ppa，eba等脑区激活更多体素，说明ccm挖掘了更多什么什么知识。
Besides, Fig.~\ref{fig:visualbrainb} (a) and (b) further illustrate the cortical activation corresponding to the semantic correlations from CCM. The results show widespread activation across high-order visual-related ROIs involved in object recognition, scene perception, and spatial-semantic integration, especially in the IPS, PPA, and OPA. This activation suggests that the decoding of semantic correlations relies on abstract and structured semantic representations in high-level visual areas, further supporting the neural plausibility of our method.

% 此外，不同细粒度的ccm特征也存在细微差异，例如，文本的粗粒度ccm在少数早期视觉皮层也有激活，而图像的细粒度ccm则几乎全部集中在高级皮层。
% 这体现了文本粗粒度ccm一定程度上也能体现图像整体的早期认知。

% In addition, different levels of CCM features exhibit subtle differences in cortical activation.
% For example, coarse-grained text CCM features in Fig.~\ref{fig:visualbrainb} (a) show activation in early visual areas (V1, V2), whereas fine-grained image CCM features are predominantly localized in high-level cortical regions. It suggests that coarse-grained text features, through semantic contextual relationship mining in the CCM, can partially decode the global early perception of vision.

% 结合cim和ccm，braincognizer能够有效补充原始pipeline在高级皮层上挖掘的语义模式，提高fmri解码信息密度，改善模态间的信息不对称。这为我们所提出的ccm和cim的有效性提供了生理解释支撑。

By integrating CIM and CCM, BrainCognizer effectively extracts more semantic patterns in high-order cortices compared with the original pipeline.
It increases the information density of fMRI decoding and mitigates inter-modal asymmetry.
These results provide physiological evidence supporting the effectiveness of the proposed CIM and CCM.

}
% \subsection{Limitations and Future Work}
% {\color{rev}Currently, our method does have certain limitations. 1) Due to significant individual variability in fMRI signals, although our approach achieves effective improvements and promising reconstruction results across multiple subjects, applying a single unified model to multiple individuals remains a challenging task. To address this issue, we plan to introduce a larger model and more data to reduce individual variability, supported by enhanced computational resources. 1) Due to the limited temporal resolution of fMRI signals, although our image reconstruction model performs well on static images, it still faces challenges when directly applied to video reconstruction tasks. To overcome this, To this end, we plan to incorporate temporal modeling methods, which offer a promising direction for reconstructing realistic videos from the brains of arbitrary individuals in the future.}

\section{Conclusion}
% This paper presents a novel fMRI-to-image reconstruction method, BrainCognizer, to improve fMRI decoding inspired by human visual cognition.
% We design the Cognitive Integration Module and Cognitive Correlation Module, which introduce diverse hierarchical semantics and regional topological relationships for better representation, to mitigate the issues of object attribute missing and relationship misalignment of elements in reconstructions.
% Moreover, we provide specific initialization for generation by Auxiliary Consistency Module to enhance the consistency of results.
% Extensive experiments demonstrate that our model has great performance and generalization ability.

To mitigate the information asymmetry between brain signals and visual stimuli, this paper proposes BrainCognizer, a novel brain decoding framework inspired by human visual cognition.
BrainCognizer incorporates the Cognitive Integration Module and the Cognitive Correlation Module, to extract hierarchical region semantics guided by human prior knowledge and capture contextual semantic relationships across regions, respectively.
They effectively enhances intra-region semantic consistency and preserves inter-region contextual relationships.
Furthermore, we analyze the interpretability of the components from a neuroscientific perspective.
Extensive experiments demonstrate that BrainCognizer achieves state-of-the-art performance across multiple evaluation metrics.

\section*{ACKNOWLEDGMENTS}
This work was supported by National Natural Science Foundation of China (grant No. 62376068, 62350710797, U24A20340), by Guangdong Basic and Applied Basic Research Foundation (grant No. 2023B1515120065), by Shenzhen Science and Technology Innovation Program (grant No. JCYJ20220818102414031).
\bibliographystyle{IEEEtran}
\bibliography{mybibliography}

% Generated by IEEEtran.bst, version: 1.14 (2015/08/26)
\begin{thebibliography}{10}
\providecommand{\url}[1]{#1}
\csname url@samestyle\endcsname
\providecommand{\newblock}{\relax}
\providecommand{\bibinfo}[2]{#2}
\providecommand{\BIBentrySTDinterwordspacing}{\spaceskip=0pt\relax}
\providecommand{\BIBentryALTinterwordstretchfactor}{4}
\providecommand{\BIBentryALTinterwordspacing}{\spaceskip=\fontdimen2\font plus
\BIBentryALTinterwordstretchfactor\fontdimen3\font minus \fontdimen4\font\relax}
\providecommand{\BIBforeignlanguage}[2]{{%
\expandafter\ifx\csname l@#1\endcsname\relax
\typeout{** WARNING: IEEEtran.bst: No hyphenation pattern has been}%
\typeout{** loaded for the language `#1'. Using the pattern for}%
\typeout{** the default language instead.}%
\else
\language=\csname l@#1\endcsname
\fi
#2}}
\providecommand{\BIBdecl}{\relax}
\BIBdecl

\bibitem{ogawa1990oxygenation}
S.~Ogawa, T.-M. Lee, A.~S. Nayak, and P.~Glynn, ``Oxygenation-sensitive contrast in magnetic resonance image of rodent brain at high magnetic fields,'' \emph{Magn Reson Med}, vol.~14, no.~1, pp. 68--78, 1990.

\bibitem{brainactiv}
D.~G. Cerdas, C.~Sartzetaki, M.~Petersen, G.~Roig, P.~Mettes, and I.~Groen, ``Brain{ACTIV}: Identifying visuo-semantic properties driving cortical selectivity using diffusion-based image manipulation,'' in \emph{ICLR}, 2025.

\bibitem{guo2025surveyfmriimagereconstruction}
W.~Guo, G.~Sun, J.~He, T.~Shao, S.~Wang, Z.~Chen, M.~Hong, Y.~Sun, and H.~Xiong, ``A survey of fmri to image reconstruction,'' \emph{arXiv:2502.16861}, 2025.

\bibitem{brainsail}
A.~Luo, J.~Yeung, R.~Zawar, S.~R. Dewan, M.~M. Henderson, L.~Wehbe, and M.~J. Tarr, ``Brain mapping with dense features: Grounding cortical semantic selectivity in natural images with vision transformers,'' in \emph{ICLR}, 2025.

\bibitem{beliy2019voxels}
R.~Beliy, G.~Gaziv, A.~Hoogi, F.~Strappini, T.~Golan, and M.~Irani, ``From voxels to pixels and back: Self-supervision in natural-image reconstruction from fmri,'' \emph{NIPS}, vol.~32, pp. 6514--6524, 2019.

\bibitem{xia2024dream}
W.~Xia, R.~De~Charette, C.~Oztireli, and J.-H. Xue, ``Dream: Visual decoding from reversing human visual system,'' in \emph{WACV}, 2024, pp. 8226--8235.

\bibitem{comby2023denoising}
P.-A. Comby, Z.~Amor, A.~Vignaud, and P.~Ciuciu, ``Denoising of fmri volumes using local low rank methods,'' in \emph{ISBI}, 2023, pp. 1--5.

\bibitem{st2018generative}
G.~St-Yves and T.~Naselaris, ``Generative adversarial networks conditioned on brain activity reconstruct seen images,'' in \emph{SMC}.\hskip 1em plus 0.5em minus 0.4em\relax IEEE, 2018, pp. 1054--1061.

\bibitem{seeliger2018generative}
K.~Seeliger, U.~G{\"u}{\c{c}}l{\"u}, L.~Ambrogioni, Y.~G{\"u}{\c{c}}l{\"u}t{\"u}rk, and M.~A. Van~Gerven, ``Generative adversarial networks for reconstructing natural images from brain activity,'' \emph{NeuroImage}, vol. 181, pp. 775--785, 2018.

\bibitem{lin2019dcnn}
Y.~Lin, J.~Li, and H.~Wang, ``Dcnn-gan: Reconstructing realistic image from fmri,'' in \emph{MVA}.\hskip 1em plus 0.5em minus 0.4em\relax IEEE, 2019, pp. 1--6.

\bibitem{ren2021reconstructing}
Z.~Ren, J.~Li, X.~Xue, X.~Li, F.~Yang, Z.~Jiao, and X.~Gao, ``Reconstructing seen image from brain activity by visually-guided cognitive representation and adversarial learning,'' \emph{NeuroImage}, vol. 228, p. 117602, 2021.

\bibitem{radford2021learning}
A.~Radford, J.~W. Kim, C.~Hallacy, A.~Ramesh, G.~Goh, S.~Agarwal, G.~Sastry, A.~Askell, P.~Mishkin, J.~Clark \emph{et~al.}, ``Learning transferable visual models from natural language supervision,'' in \emph{ICML}, 2021, pp. 8748--8763.

\bibitem{ddpm}
J.~Ho, A.~Jain, and P.~Abbeel, ``Denoising diffusion probabilistic models,'' \emph{NIPS}, vol.~33, pp. 6840--6851, 2020.

\bibitem{ldm}
R.~Rombach, A.~Blattmann, D.~Lorenz, P.~Esser, and B.~Ommer, ``High-resolution image synthesis with latent diffusion models,'' in \emph{CVPR}, 2022, pp. 10\,684--10\,695.

\bibitem{vd}
X.~Xu, Z.~Wang, G.~Zhang, K.~Wang, and H.~Shi, ``Versatile diffusion: Text, images and variations all in one diffusion model,'' in \emph{ICCV}, 2023, pp. 7754--7765.

\bibitem{ozcelik2023natural}
F.~Ozcelik and R.~VanRullen, ``Natural scene reconstruction from fmri signals using generative latent diffusion,'' \emph{Sci. Rep.}, vol.~13, no.~1, p. 15666, 2023.

\bibitem{scotti2024reconstructing}
P.~Scotti, A.~Banerjee, J.~Goode, S.~Shabalin, A.~Nguyen, A.~Dempster, N.~Verlinde, E.~Yundler, D.~Weisberg, K.~Norman \emph{et~al.}, ``Reconstructing the mind's eye: fmri-to-image with contrastive learning and diffusion priors,'' \emph{NIPS}, vol.~36, 2024.

\bibitem{takagi2023high}
Y.~Takagi and S.~Nishimoto, ``High-resolution image reconstruction with latent diffusion models from human brain activity,'' in \emph{CVPR}, 2023, pp. 14\,453--14\,463.

\bibitem{wang2024mindbridge}
S.~Wang, S.~Liu, Z.~Tan, and X.~Wang, ``Mindbridge: A cross-subject brain decoding framework,'' in \emph{CVPR}, 2024, pp. 11\,333--11\,342.

\bibitem{luanimate}
Y.~Lu, C.~Du, C.~Wang, X.~Zhu, L.~Jiang, X.~Li, and H.~He, ``Animate your thoughts: Reconstruction of dynamic natural vision from human brain activity,'' in \emph{ICLR}, 2025.

\bibitem{felleman1991distributed}
D.~J. Felleman and D.~C. Van~Essen, ``Distributed hierarchical processing in the primate cerebral cortex.'' \emph{Cerebral cortex (New York, NY: 1991)}, vol.~1, no.~1, pp. 1--47, 1991.

\bibitem{grill2014functional}
K.~Grill-Spector and K.~S. Weiner, ``The functional architecture of the ventral temporal cortex and its role in categorization,'' \emph{Nat. Rev. Neurosci.}, vol.~15, no.~8, pp. 536--548, 2014.

\bibitem{kravitz2011new}
D.~J. Kravitz, K.~S. Saleem, C.~I. Baker, and M.~Mishkin, ``A new neural framework for visuospatial processing,'' \emph{Nat. Rev. Neurosci.}, vol.~12, no.~4, pp. 217--230, 2011.

\bibitem{huth2012continuous}
A.~G. Huth, S.~Nishimoto, A.~T. Vu, and J.~L. Gallant, ``A continuous semantic space describes the representation of thousands of object and action categories across the human brain,'' \emph{Neuron}, vol.~76, no.~6, pp. 1210--1224, 2012.

\bibitem{lu2023minddiffuser}
Y.~Lu, C.~Du, Q.~Zhou, D.~Wang, and H.~He, ``Minddiffuser: Controlled image reconstruction from human brain activity with semantic and structural diffusion,'' in \emph{ACM MM}, 2023, pp. 5899--5908.

\bibitem{koffka1922perception}
K.~Koffka, ``Perception: an introduction to the gestalt-theorie.'' \emph{Psychol. Bull.}, vol.~19, no.~10, p. 531, 1922.

\bibitem{masland2012neuronal}
R.~H. Masland, ``The neuronal organization of the retina,'' \emph{Neuron}, vol.~76, no.~2, pp. 266--280, 2012.

\bibitem{poggio1990network}
T.~Poggio and S.~Edelman, ``A network that learns to recognize three-dimensional objects,'' \emph{Nature}, vol. 343, no. 6255, pp. 263--266, 1990.

\bibitem{brainclip}
Y.~Liu, Y.~Ma, W.~Zhou, G.~Zhu, and N.~Zheng, ``Brainclip: Bridging brain and visual-linguistic representation via clip for generic natural visual stimulus decoding,'' \emph{arXiv:2302.12971}, 2023.

\bibitem{lin2022mind}
S.~Lin, T.~Sprague, and A.~K. Singh, ``Mind reader: Reconstructing complex images from brain activities,'' \emph{NIPS}, vol.~35, pp. 29\,624--29\,636, 2022.

\bibitem{rpi}
T.~Fang, Y.~Qi, and G.~Pan, ``Reconstructing perceptive images from brain activity by shape-semantic gan,'' \emph{NIPS}, vol.~33, pp. 13\,038--13\,048, 2020.

\bibitem{ozcelik2022reconstruction}
F.~Ozcelik, B.~Choksi, M.~Mozafari, L.~Reddy, and R.~VanRullen, ``Reconstruction of perceived images from fmri patterns and semantic brain exploration using instance-conditioned gans,'' in \emph{IJCNN}.\hskip 1em plus 0.5em minus 0.4em\relax IEEE, 2022, pp. 1--8.

\bibitem{gu2024decoding}
Z.~Gu, K.~Jamison, A.~Kuceyeski, and M.~R. Sabuncu, ``Decoding natural image stimuli from f{MRI} data with a surface-based convolutional network,'' in \emph{MIDL}, 2024, pp. 107--118.

\bibitem{yang2024brain}
H.~Yang, J.~Gee, and J.~Shi, ``Brain decodes deep nets,'' in \emph{CVPR}, 2024, pp. 23\,030--23\,040.

\bibitem{wang2023better}
A.~Y. Wang, K.~Kay, T.~Naselaris, M.~J. Tarr, and L.~Wehbe, ``Better models of human high-level visual cortex emerge from natural language supervision with a large and diverse dataset,'' \emph{Nat. Mach. Intell}, vol.~5, no.~12, pp. 1415--1426, 2023.

\bibitem{zhouclip}
Q.~Zhou, C.~Du, S.~Wang, and H.~He, ``Clip-mused: Clip-guided multi-subject visual neural information semantic decoding,'' in \emph{ICLR}, 2024.

\bibitem{vae}
D.~P. Kingma and M.~Welling, ``Auto-encoding variational bayes,'' \emph{arXiv preprint arXiv:1312.6114}, 2013.

\bibitem{kmeans}
J.~MacQueen, ``Some methods for classification and analysis of multivariate observations,'' in \emph{Berkeley Symp. on Math. Statist. and Prob.}, vol.~5, 1967, pp. 281--298.

\bibitem{allen2022massive}
E.~J. Allen, G.~St-Yves, Y.~Wu, J.~L. Breedlove, J.~S. Prince, L.~T. Dowdle, M.~Nau, B.~Caron, F.~Pestilli, I.~Charest \emph{et~al.}, ``A massive 7t fmri dataset to bridge cognitive neuroscience and artificial intelligence,'' \emph{Nat. Neurosci}, vol.~25, no.~1, pp. 116--126, 2022.

\bibitem{han2024mindformer}
I.~Han, J.~Lee, and J.~C. Ye, ``Mindformer: A transformer architecture for multi-subject brain decoding via fmri,'' \emph{arXiv:2405.17720}, 2024.

\bibitem{coco}
T.-Y. Lin, M.~Maire, S.~Belongie, J.~Hays, P.~Perona, D.~Ramanan, P.~Doll{\'a}r, and C.~L. Zitnick, ``Microsoft coco: Common objects in context,'' in \emph{ECCV}, 2014, pp. 740--755.

\bibitem{wang2004image}
Z.~Wang, A.~C. Bovik, H.~R. Sheikh, and E.~P. Simoncelli, ``Image quality assessment: from error visibility to structural similarity,'' \emph{TIP}, vol.~13, no.~4, pp. 600--612, 2004.

\bibitem{krizhevsky2012imagenet}
A.~Krizhevsky, I.~Sutskever, and G.~E. Hinton, ``Imagenet classification with deep convolutional neural networks,'' \emph{NIPS}, vol.~25, pp. 1106--1114, 2012.

\bibitem{szegedy2016rethinking}
C.~Szegedy, V.~Vanhoucke, S.~Ioffe, J.~Shlens, and Z.~Wojna, ``Rethinking the inception architecture for computer vision,'' in \emph{CVPR}, 2016, pp. 2818--2826.

\bibitem{tan2019efficientnet}
M.~Tan and Q.~Le, ``Efficientnet: Rethinking model scaling for convolutional neural networks,'' in \emph{ICML}, 2019, pp. 6105--6114.

\bibitem{caron2020unsupervised}
M.~Caron, I.~Misra, J.~Mairal, P.~Goyal, P.~Bojanowski, and A.~Joulin, ``Unsupervised learning of visual features by contrasting cluster assignments,'' \emph{NIPS}, vol.~33, pp. 9912--9924, 2020.

\bibitem{mcinnes2018umap}
L.~McInnes, J.~Healy, and J.~Melville, ``Umap: Uniform manifold approximation and projection for dimension reduction,'' \emph{arXiv:1802.03426}, 2018.

\bibitem{neuropictor}
J.~Huo, Y.~Wang, Y.~Wang, X.~Qian, C.~Li, Y.~Fu, and J.~Feng, ``Neuropictor: Refining fmri-to-image reconstruction via multi-individual pretraining and multi-level modulation,'' in \emph{ECCV}, 2024, pp. 56--73.

\end{thebibliography}
% \end{CJK}

\end{document}